\documentstyle[preprint,aps]{revtex}
\tightenlines

\begin{document}
\title{{\bf EFFECTS OF EPITAXIAL STRAIN ON THE GROWTH MECHANISM OF YBa}$_{2}${\bf Cu%
}$_{3}${\bf O}$_{7-x}${\bf \ THIN FILMS IN [YBa}$_{2}${\bf Cu}$_{3}${\bf O}$%
_{7-x}/${\bf PrBa}$_{2}${\bf Cu}$_{3}${\bf O}$_{7-x}${\bf ] SUPERLATTICES}}
\author{M. Varela$^{a),c)}$, W. Grogger$^{b)}$, D. Arias$^{a),\S }$, Z. Sefrioui$%
^{a)}$, C. Le\'{o}n$^{a)}$, L. Vazquez$^{d)}$, C. Ballesteros$^{c)}$, K. M.
Krishnan$^{b)}$, J. Santamar\'{i}a$^{a)}$}
\address{$^{a)}${\it \ GFMC, Dpto F\'{i}sica Aplicada III, Universidad Complutense, }%
28040 Madrid. Spain\\
$^{b)}${\it \ Materials Sciences Division, 72-222, Lawrence Berkeley }%
National Laboratory, University of California, Berkeley, CA 94720. U.S.A.\\
$^{c)}${\it \ Dpto F\'{i}sica, Universidad Carlos III de Madrid, } 28911\\
Legan\'{e}s, Madrid. Spain \\
$^{d)}${\it \ Instituto de Ciencia de Materiales ICMM-CSIC, }28049 Madrid.\\
Spain\\
$^{\S }$On leave from Universidad del Quindio. Armenia. Colombia}
\date{February 27$^{th}$ 2002}
\maketitle
\pacs{74.76.Bz, 74.80.Dm, 68.65.Cd}

\begin{abstract}
We report on the growth mechanism of YBa$_{2}$Cu$_{3}$O$_{7-x}$. Our study
is based on the analysis of ultrathin, YBa$_{2}$Cu$_{3}$O$_{7-x}$ layers in
c-axis oriented YBa$_{2}$Cu$_{3}$O$_{7-x}$ / PrBa$_{2}$Cu$_{3}$O$_{7}$
superlattices. We have found that the release of epitaxial strain in very
thin YBCO layers triggers a change in the dimensionality of the growth mode.
Ultrathin, epitaxially strained, YBCO layers with thickness below 3 unit
cells grow in a block by block two dimensional mode coherent over large
lateral distances. Meanwhile, when thickness increases, and the strain
relaxes, layer growth turns into three dimensional, resulting in rougher
layers and interfaces.
\end{abstract}

\section*{INTRODUCTION\protect\bigskip}

In recent years, thin films of complex ionic oxides have played a
significant role in the discovery and understanding of important new
phenomena in condensed matter physics. Basic studies or important related
applications in various fields like high Tc superconductivity, colossal
magnetoresistance, etc, require the growth of ultrathin films with
controlled composition and morphology. In this respect, the specific growth
mode seriously influences the final film structure and morphology, and thus,
can deeply condition its physical properties. Growth mechanisms are
determined by the chemical composition, type of chemical bonding, etc, and
therefore they may be intrinsically different when going from semiconductors
to metals or ionic oxides. While the questions of which is the growth
mechanism and which is the minimum growth unit, have been satisfactory
addressed in semiconductors or metals, in the case of ionic oxides, the
growth process is still not fully understood due, probably, to the
complexity of the unit cell. Experiments directed to the analysis or
application of the physical properties of those ultrathin layers require an
accurate control of the growth process at atomic scale. In fact, many
studies fail because the chemical and physical disorder at the interfaces
(interdiffusion and roughness) determined by the particular growth mechanism
may invalidate the conclusions obtained for ultrathin layers [1,2].

\bigskip

It has been reported that in the initial stages, the growth of c-axis
oriented YBa$_{2}$Cu$_{3}$O$_{7-x}$ (YBCO) films takes place quite smoothly,
in a two-dimensional (2D) mode [3-5]. From the analysis of superlattices
with a non-integer number of YBa$_{2}$Cu$_{3}$O$_{7-x}$ cells, in a recent
letter we have reported evidence for a two-dimensional block-by-block growth
mechanism in which the growth units are single unit cells [6]. We use the
term block-by-block in the sense that the high substrate temperature (900 
%TCIMACRO{\UNICODE{0xba}}%
%BeginExpansion
${{}^o}$%
%EndExpansion
C) provides the necessary mass transport for complete unit cell blocks to
nucleate. We showed that it is possible to obtain laterally coherent one
unit cell thick layers. On the other hand a three- dimensional growth
mechanism around screw dislocations has been reported for thicker films,
leading to a rough terrace-like surface morphology [7,8]. In fact this 3D
growth has been recently shown to determine the critical current of YBCO
thin films. Linear defects (edge and screw dislocations) associated to a 3D
growth provide the strong pinning centres responsible for enhanced critical
currents [7,8].

\bigskip

Since the growth of YBCO has been reported to take place in a 2D mode for
very thin films, and in a 3D mode for thicker films, the question here is
what is the driving mechanism to change the growth mode from 2D to 3D. In
this paper we present a consistent approach based on ``ex-situ''
quantitative structural and chemical characterization of epitaxially grown
YBa$_{2}$Cu$_{3}$O$_{7-x}$ (YBCO) layers in c-axis oriented YBa$_{2}$Cu$_{3}$%
O$_{7-x}$/PrBa$_{2}$Cu$_{3}$O$_{7} $(YBCO/PBCO) superlattices. The
superlattice structure is of particular interest because interface planes
between YBCO and PBCO were once the growth fronts, and therefore they
represent an important information source to study the growth process
itself. The study of the growth modes of ultrathin layers requires a
quantitative structural and chemical characterization at relevant length
scales. Roughness has been analyzed by complementary reciprocal and real
space techniques, x-ray diffraction (XRD) and energy filtered transmission
electron microscopy (EFTEM). Combining both techniques it is possible to
accurately quantify the interfacial roughness in both the growth and lateral
directions [9]. We have found a correlation between the release of epitaxial
strain in YBCO layers, when a certain critical thickness is attained, and
the appearance of roughness in the superlattices, which increases with layer
thickness and confirms a 3D-like growth in relaxed samples.

\bigskip

\section*{EXPERIMENTAL\protect\bigskip}

High quality YBCO/PBCO superlattices grown by high pressure sputtering (3.6
mbar pure oxygen)on SrTiO$_{3}$ substrates at 900%
%TCIMACRO{\UNICODE{0xba}}%
%BeginExpansion
${{}^o}$%
%EndExpansion
C. This technique provides a very thermalized and ordered growth at a slow
rate (0.013 nm/s for YBCO), which allows a very accurate control of layer
thickness [10]. The films for this study were [YBCO$_{n}$/PBCO$_{5}$]$%
_{1000\AA }$ superlattices with nominal values for YBCO layer thickness n
comprised between 1 and 12 unit cells up to a total thickness of 1000 \AA . $%
\theta $-2$\theta $ x-ray spectra were obtained in a conventional
diffractometer using CuK$\alpha $ radiation. High angle XRD spectra were
analysed using the SUPREX9 software, which allows obtaining quantitative
values for interface disorder parameters like interdiffusion or roughness
[11]. Energy-Filtered Transmission Electron Microscopy (EFTEM)
investigations were performed on cross-section samples on a Philips
CM200/FEG equipped with a field emission source and a post-column
energy-filter (Gatan Imaging Filter, GIF). For imaging we used the low
energy edges of Y and Pr, applying the two-window technique (jump ratio
images, i.e., division of an image located right on the ionization edge by
an image acquired before the edge onset) [12-15]. In this manner individual
unit cells of YBCO and PBCO in the growth direction were resolved, and
individual atomic planes of Y and Pr were detected.

\bigskip

\section*{RESULTS AND DISCUSSION}

\bigskip

A set of element-specific energy filtered images, using the characteristic
Pr.N$_{4,5}$ionization edge, is shown in figure 1. The darkest contrast
corresponds to YBCO layers, while the lightest contrast corresponds to PBCO
layers. Energy filtered images using the characteristic Y-M $_{4,5}$ energy
edge give complementary images to the ones obtained with Pr mapping [6].The
images correspond to a [YBCO$_{2}$/PBCO$_{5}$]$_{1000\AA }$ (fig 1(a)) and a
[YBCO$_{8}$/PBCO$_{5}$]$_{1000\AA }$ (fig 1(b)) superlattices. Magnification
of the images is low, in order to get enough structural information over
wide lateral scales, larger than sample thickness. While for the ultrathin
YBCO layers (fig 1(a)) perfectly smooth layers of uniform thickness can be
observed over long lateral distances, clear layer undulations can be
observed in the thicker YBCO films of the [YBCO$_{8}$/PBCO$_{5}$]
superlattice (fig1(b)), resulting from thickness fluctuations with lateral
length scales typically comparable to sample thickness. These images can be
statistically treated to quantify roughness. Images have been digitized to
find the locii of the midpoint of Y in each bilayer thus providing a lateral
height profile of the individual bilayers. The standard deviation of the
height distribution for each bilayer (bilayer roughness $\sigma _{bilayer}$
hereafter) can be used to examine roughness evolution along the growth
direction. Alternatively the height differences between Y profiles for
consecutive bilayers at the same lateral coordinate provide a map of the
fluctuations in the modulation length, $\Lambda $, with $\Lambda $=N$_{A}$c$%
_{A}$+N$_{B}$c$_{B}$, where c$_{A}$, c$_{B}$, N$_{A}$, and N$_{B}$ are the
lattice parameters and number of unit cells of materials A and B,
respectively. The standard deviation of the distribution of modulation
lengths, $\sigma $, can be used as a measure of step disorder roughness in
superlattices. Obviously the two quantities $\sigma $ and $\sigma _{bilayer}$
will be different if roughness is vertically correlated. This method has
been proposed recently and satisfactory applied to analyse interface
roughness in Fe/Cr superlattices [9]. Histogram plots of the modulation
length distributions are shown in figures 2(a) and 2(b) respectively for the
[YBCO$_{2}$/PBCO$_{5}$]$_{1000\AA }$ and [YBCO$_{8}$/PBCO$_{5}$]$_{1000\AA }$
superlattices of figure 1. As depicted in 2(a) for the [YBCO$_{2}$/PBCO$_{5}$%
]$_{1000\AA }$ superlattice a quite sharp-peaked distribution centered in
81\AA\ is obtained for $\Lambda $, with a standard deviation of less than 1
\AA . However, the same analysis performed on the [YBCO$_{8}$/PBCO$_{5}$]$%
_{1000\AA }$ sample, in figure 2(b) , shows a gaussian-like distribution of
data around a mean value of 151 \AA\ with a standard deviation of 6 \AA . It
is important to note that, the average value of the modulation length over
the superlattice stacking was in agreement with the nominal overall
composition of the superlattice. This shows that thickness fluctuations
occur locally as a consequence of the growth process and do not result of
uncontrolled fluctuations in the deposition rate. Roughness of each bilayer $%
\sigma _{bilayer}$ has been obtained from the standard deviation of the
height distribution for each bilayer in the [YBCO$_{8}$/PBCO$_{5}$]$%
_{1000\AA }$ superlattice. Figure 3 shows the bilayer roughness as a
function of the bilayer index in the stacking. From the observation of
figure 3 one can see how the roughness increases cumulatively with bilayer
index. The dotted line in figure 3 is a fit to a power law of the form, $%
\sigma _{bilayer}$ =$\sigma _{a}$ N$^{\alpha }$, being $\sigma _{a}$ = 5.2
\AA\ the roughness in the first bilayer, $\alpha $ =0.51 an exponent which
accounts for the cumulative roughness evolution, and N is the bilayer index
[9,16]. No fluctuations were found in the height distribution of each
bilayer in the [YBCO$_{2}$/PBCO$_{5}$]$_{1000\AA }$ superlattice within the
lateral size of the images.

\bigskip

Figure 4 shows the XRD spectra of a [YBCO$_{1}$/PBCO$_{5}$] sample (bottom)
and a [YBCO$_{8}$/PBCO$_{5}$] superlattice (top), displaced vertically for
clarity. XRD spectra of the [YBCO$_{1}$/PBCO$_{5}$] sample show sharp
superlattice satellite peaks around the main Bragg peaks, indicating a good
structural quality. X ray data were analysed with the SUPREX code [11].
Lines in figure 4 are the refinement calculations supplied by the SUPREX
program. The structural model used in the refinement calculation refines
thicknesses of the individual layers, c-lattice parameters and interface
disorder related parameters. Interface roughness is accounted for by
allowing the bilayer thickness (modulation length) to fluctuate around a
mean value in a gaussian way [11]. Interface roughness compares with the
standard deviation ($\sigma $) of the gaussian layer thickness distribution
obtained from the EFTEM analysis. The analysis of the XRD spectra of samples
with thin (1-3 unit cells) YBCO layers did not show measurable interface
roughness. On the other hand, YBCO layers with thickness above 4 unit cells,
show clear damping of the high order satellite peaks as a result of enhanced
structural disorder, (see figure 4). In this case the structural refinement
confirms the presence of random layer thickness fluctuations which increases
with YBCO layer thickness (see figure 5(a)). Good agreement was found
between roughness values obtained from XRD and from electron microscopy. The
roughness parameter $\sigma $=8 \AA\ obtained from x-rays for the [YBCO$_{8}$%
/PBCO$_{5}$] superlattice has to be compared with the standard deviation of
the distribution of modulation lengths (6 \AA ) obtained from the electron
microscopy analysis. We have previously shown that superlattices with YBCO
layer thickness below 4 unit cells grow epitaxially strained due to lattice
mismatch to PBCO [10,17]. Figure 5 shows the evolution of the YBCO c-lattice
parameter with YBCO layer thickness. It can be observed that for YBCO
thickness below 4 unit cells there is a significant contraction of the c
lattice spacing arising from the 1\% lattice mismatch to PBCO according to
Poisson effect [17]. Interestingly, roughness increases sharply above 3 unit
cells, when epitaxial strain relaxes (see figure 5). When increasing YBCO
thickness over this critical value the strain relaxes, and simultaneously
the superlattice interfaces become rougher. These results point to a change
in the growth mechanism of YBCO from a 2D-like in epitaxially strained
samples to a 3D rough growth for YBCO thickness over 4 unit cells. An
interesting question is whether the 3D growth mode of the rough relaxed
samples already provides the linearly correlated disorder (edge and screw
dislocations) which have recently shown to effectively pin vortices in
thicker (140 nm) samples [7].We have done measurements of the angular
dependence of the magnetoresistance at temperatures ranging between 0.7 T$_{%
{\bf c}}$ and 0.99 T$_{{\bf c}}$, in magnetic fields ranging from 100 to
90000 gauss. No evidence was found for ''cusp''- like features
characteristic of vortex confinement by correlated disorder. This evidences
that thicker samples are necessary for these defects to develop and /or to
be effective for vortex pinning.

\bigskip

Attempts were done to artificially induce roughness in the {\it ultrathin}
YBCO layers growing a non integer number of YBCO unit cells. Figure 6 shows
an element-specific energy-filtered images, using the characteristic Pr-N$%
_{4,5}$ edge for a [YBCO$_{1.5}$/PBCO$_{5}$] superlattice. Lattice fringes
are observed, though this contrast is not element specific. The
superlattice, with a non-integer number of unit cells of YBCO in the stack,
shows abrupt changes in the YBCO layer thickness from one to two unit cells,
through steps one unit cell high, over a lateral length scale ranging from
100-200 nm. The specimen thickness in the direction parallel to the electron
beam is of the order of 10-30 nm, well below the lateral dimension where a
film of uniform thickness can be grown. In this incommensurate
superlattices, layer discontinuities and interface steps are observed,
leading to non-smooth interfaces, but the spatial distribution of such
defects seems to be non random.

\bigskip

Low angle (2$\theta \leq $10%
%TCIMACRO{\UNICODE{0xba}}%
%BeginExpansion
${{}^o}$%
%EndExpansion
) x-ray diffraction spectra of a set of [YBCO$_{n}$/PBCO$_{5}$]
superlattices with n ranging between 1 and 2 YBCO unit cells are shown in
figure 7. In this angular range, x rays are not sensitive to crystal
structure, but to the chemical modulation, and due to the x-ray grazing
incidence the spectra provide information averaged over large areas of the
sample. From bottom to top the spectra depicted correspond to n=1.2, 1.5 and
2 YBCO unit cells. Finite size oscillations corresponding to sample
thickness can be detected, denoting a flat sample surface. Low angle
superlattice peaks, labeled by m, and sharp satellite peaks, corresponding
to the first superlattice Bragg peak (denoted by n) are also observed. Low
angle peaks are determined by the modulation length, $\Lambda $.
Superlattice Bragg peaks are determined by the average lattice spacing c= $%
\Lambda $/(N$_{A}$+N$_{B}$). Satellite peaks corresponding to the first
Bragg peak will occur for q vectors given by q=2$\pi $ /c - 2$\pi ${\it l}/$%
\Lambda $, where {\it l} are integers. Thus, if the modulation length is a
non-integer number times the average lattice spacing, low angle peaks and
satellites from the first Bragg peak will not merge [18]. Moreover, in the
presence of step-disorder the non-integer modulation arises from a layer
thickness fluctuation around a mean value, with a standard deviation
quantifying the step-disorder roughness. This randomness causes satellite
broadening and overshadows the apparent splitting in most cases. However,
the analysis of the XRD patterns for integer number of YBCO layers has shown
the absence of step-disorder in our superlattices [10]. Hence, in the
absence of step-disorder, the layer thickness does not fluctuate in a random
fashion, and an unusual superlattice diffraction pattern arises with sharp
diffraction satellite peaks. Such a splitting in satellite orders is
noticeable in figure 7 confirming that composition changes coherently from
interface plane to interface plane. A similar behaviour has been previously
observed in MBE grown semiconductor superlattices [18] and also in complex
oxide superlattices [6,19]. These x-ray scattering results provide strong
evidence for a lateral growth mode in which the growth units are complete
YBCO units cells, i.e. a 2D block-by-block growth mechanism and layer growth
would take place by the lateral deposition of blocks one unit cell in height
[6,20]. The composition of the interface plane is observed to change
coherently from one bilayer to the next one as a result of a coherent 2D
layer growth [18]. The coherent 2D growth of strained layers minimize the
interface energy at the cost of the excess of elastic energy due to lattice
mismatch. Increasing YBCO thickness relaxes stored elastic energy, and
triggers a change to a 3D growth mode which results in terrace-like and
spiral growth. Figure 8 shows an atomic force microscopy (AFM) image of a
thick (300 \AA ) YBCO film on a 5 u.c. PBCO buffer, showing a distribution
of pyramid structures. Terraces are 250 nm wide and steps between terraces
of the height of one u.c. are observed. This clearly shows that a 3D growth
mechanism is developed upon strain relaxation.

Finally, it is interesting to compare the results obtained for the growth of
YBCO in YBCO/PBCO superlattices with the growth of single YBCO layers
[21,22]. Thin YBCO layers were grown on single STO substrates (100)
oriented. Layer thickness was determined from low angle x-ray reflectivity
oscillations. For YBCO layer thickness below 10 nm a significant c lattice
contraction is observed (see figure 9 (a)), which correlates with a
substantial reduction of the critical temperature (see figure 9 (b)).
Critical temperatures are evaluated at zero resistivity. It is important to
note that critical temperatures in excess of 80 K are practically recovered
for YBCO thickness larger than 10 nm. This results are consistent with
recent calculations which propose a critical thickness of YBCO on STO of 80
nm [23]. Note that despite a larger lattice mismatch with STO (-1.4\% and
-0.7 \% along a and b axis respectively) than with PBCO (-0.7 \% and -1.1 \%
along a and b axis respectively) the critical thickness is larger for films
grown on STO. This can be discussed from two different viewpoints.On the one
hand the superlattice geometry in which the YBCO layers are sandwiched
between PBCO layers stresses YBCO more efficiently. This method would
probably underestimate the critical thickness of single films grown on PBCO
buffer layers. On the other hand, and probably more important, ultrathin
layers are very delicate and exposure to ambient atmosphere can chemically
affect the surface and degrade the superconducting properties. We want to
remark that these results may also depend on growth parameters and
deposition technique. Substantially different results have been recently
reported [8] for films of comparable thickness grown by pulsed laser
deposition (PLD) at significanly lower growth temperatures (770-850 
%TCIMACRO{\UNICODE{0xba}}%
%BeginExpansion
${{}^o}$%
%EndExpansion
C) than in this work (900 
%TCIMACRO{\UNICODE{0xba}}%
%BeginExpansion
${{}^o}$%
%EndExpansion
C). Reference [8] shows an increase of the c-lattice parameter when the YBCO
is reduced, contrary to the expectations of Poisson effect.

\bigskip

In summary, we have shown that strained ultrathin YBCO layers on PBCO grow
in a two-dimensional block-by-block mode, where the minimum growth unit is a
whole oxide unit cell. Layer grows laterally by the successive deposition of
one unit cell blocks ``wetting'' the surface. When increasing YBCO layer
thickness over 3-4 unit cells epitaxial strain relaxes, and random interface
roughness arises which increases cumulatively with YBCO layer thickness.
Strain relaxation triggers a change in the growth mode from 2D to 3D. This
3D growth mode in thick YBCO layers is in agreement with previous
observations of growth pyramids, with steps one unit cell high, in thick
YBCO layers.

\section{\newpage ACKNOWLEDGMENTS}

We thank Prof. Ivan Schuller for helpful conversations and interesting
suggestions. Work supported by spanish CICYT MAT 2000 1468. Work at
LBNL/NCEM was supported by the Director, Office of Energy Research, Office
of Basic Energy Sciences, Materials Sciences Division of the U.S. Department
of Energy under contract No. DE-AC03-76SF00098.\bigskip \bigskip

\bigskip

\section*{REFERENCES}

\bigskip

[1] E. E. Fullerton, J. Guimpel, O. Nakamura, I. K. Schuller, Phys. Rev.
Lett {\bf 69}, 2859 (1992).

\bigskip

[2] I. N. Chan, D.C. Vier, O. Nakamura, J. Hasen, J. Guimpel, S. Schultz ,
I.K. Schuller, Phys. Lett. A {\bf 175}, 241 (1993).

\bigskip

[3] C. L. Jia, H. Soltner, G. Jacob, T. Hahn, H. Adrian, K. Urban. Physica 
{\bf C 210}, 1 (1993).

\bigskip

[4] S. J. Pennycook, M. F. Chisholm, D. E. Jesson, D. P. Norton, D. H.
Lowndes, R. Feenstra, and H. R. Kerchner, J. O. Thomson. Phys. Rev. Lett. 
{\bf 67}, 765 (1991).

\bigskip

[5] T. Terashima, Y. Bando, K. Iijima, K. Yamamoto, K. Hirata, K. Hayashi,
K. Kamigaki, and H. Terauchi. Phys. Rev. Lett. {\bf 67}, 1362 (1991)

\bigskip

[6] M. Varela, W. Grogger, D. Arias, Z. Sefrioui, C. Le\'{o}n, C.
Ballesteros, K. M. Krishnan, J. Santamar\'{i}a. Phys. Rev. Lett. {\bf 86},
5156 (2001).

\bigskip

[7] B. Dam, J. M. Huijbregtse, F. C. KLaassenk, R. C. F. Van der Geest, G.
Doornbos,J. H. Rector, A. M. Testa, S. Freisem, J. C. Martinez, B.
St\"{a}uble-P\"{u}mpin \&R. Griessen, Nature {\bf 399}, 439 (1999)

\bigskip

[8] B. Dam, J. M. Huijbregtse, and J. H. Rector, Phys. Rev. B {\bf 65},
064528 (2002)

\bigskip

[9] M.C. Cyrille, S. Kim, M.E. G\'{o}mez, J. Santamar\'{i}a, K.M. Krishnan
and I.K. Schuller. Phys. Rev. B {\bf 62}, 3361 (2000)

\bigskip

[10] M. Varela, D. Arias, Z. Sefrioui, C. Le\'{o}n, C. Ballesteros, J.
Santamar\'{i}a. Phys. Rev. B. {\bf 62} 12509 (2000).

\bigskip

[11] E. E. Fullerton, I. K. Schuller, H. Vanderstraeten, Y. Bruynseraede,
Phys. Rev. B {\bf 45}, 9292 (1992).

\bigskip

[12] O. L. Krivanek, M. K. Kundmann, K. Kimoto, J. Microsc. {\bf 180} 277
(1995).

\bigskip

[13] F. Hofer, W. Grogger, G. Kothleitner, P. Warbichler, Ultramicroscopy 
{\bf 67}, 83 (1997).

\bigskip

[14] T. Navidi-Kasmai, H. Kohl, Ultramicroscopy {\bf 81}, 223 (2000).

\bigskip

[15] O. L. Krivanek, A. J. Gubbens, M. K. Kundmann, G. C. Carpenter, Proc.
51st Annual Meeting Microsc. Soc. Am., eds. G. W. Bailey $\backslash $\& C.
L. Rieder (San Francisco Press, San Francisco), 586 (1993).

\bigskip

[16] A.L. Barabasi, H.E. Stanley. Fractal concepts in surface growth.
Cambridge Univ. Press (1995)

\bigskip

[17] M. Varela, Z. Sefrioui , D. Arias, M. A. Navacerrada, M. Luc\'{i}a, M.
A. L\'{o}pez de la Torre, C. Le\'{o}n, G. Loos, F. S\'{a}nchez-Quesada and
J. Santamar\'{i}a. Phys. Rev. Lett. {\bf 83} 3936 (1999)

\bigskip

[18] I. K. Schuller, M. Grimsditch, F. Chambers, G. Devane, H.
Vanderstraeten, D. Neerinck, J.-P. Locquet, and Y. Bruynseraede, Phys. Rev.
Lett. {\bf 65}, 1235 (1990).

\bigskip

[19] G. Balestrino, G. Pasquini, A. Tebano. Phys.Rev.B {\bf 62} 1421 (2000)

\bigskip

[20] J. P. Locquet, A. Catana, E. M\"{a}chler, C. Gerber, J. G. Bednorz,
Appl. Phys. Lett. {\bf 64}, 372 (1994).

\bigskip

[21] L. Cao and J. Zegenhagen, Phys. Stat. Sol. (b) {\bf 215}, 587 (1999)

\bigskip

[22] L. X. Cao, J. Zegenhagen, M. Cardona, C. Giannini, L. De Caro, and L.
Tapfer, J. Appl. Phys. {\bf 91}, 1265 (2002)

\bigskip

[23] L. X. Cao, T. L. Lee, F. Renner, Y. Su, R. L. Johnson, and J.
Zegenhagen, Phys. Rev. {\bf B 65}, 113402 (2002)

\bigskip

\bigskip

\section{\newpage}

\section{\protect\bigskip FIGURE CAPTIONS}

Figure 1: (a) Pr-N $_{4,5}$ energy filtered image of a [YBCO$_{2}$/PBCO$_{5}$%
] superlattice in cross section geometry. White contrast corresponds to PBCO
layers, while YBCO layers appear dark. (b) Pr-N4,5 energy filtered image of
a [YBCO$_{8}$/PBCO$_{5}$] superlattice. Note the fluctuations in layer
thicknesses.

\bigskip

Figure 2: (a) Histogram plot of the distribution of modulation lengths $%
\Lambda $ obtained from the EFTEM images in the [YBCO$_{2}$ /PBCO$_{5}$ ]
superlattice. The dotted line is a gaussian fit, with a standard deviation
of 0.7 \AA\ (b) Histogram plot corresponding to the [YBCO$_{8}$/PBCO$_{5}$ ]
superlattice. The gaussian curve fit shows a standard deviation of 6 \AA .

Figure 3: Roughness of each bilayer versus bilayer index, N, for the [YBCO$%
_{8}$/PBCO$_{5}$ ] sample. The dotted line is a fit to a power law $\sigma $=%
$\sigma _{a}$ N$^{\alpha }$, being $\sigma _{a}$ = 5.2 \AA\ the roughness in
the first bilayer and $\alpha $ =0.51 an exponent which accounts for the
roughness evolution.

\bigskip

Figure 4: XRD spectra (dots) together with the structural refinement
(superimposed solid line) for a [YBCO$_{1}$/PBCO$_{5}$] superlattice
(bottom) and a [YBCO$_{8}$/PBCO$_{5}$] sample (top). Note how high order
satellites are damped in the second case.

\bigskip

Figure 5: (a) Step disorder roughness ($\sigma $) extracted form the XRD
refinement versus YBCO layer thickness for a set of [YBCO$_{n}$/PBCO$_{5}$]
superlattices with 12$\succ $n$\succ $1 u.c. (b) Evolution of the YBCO
c-lattice parameter with YBCO layer thickness for the same samples. When
YBCO thickness increases, epitaxial strain relaxes and roughness shows up.

\bigskip

Figure 6: (a) Element-specific energy-filtered images, using the
characteristic Pr-N $_{4,5}$ edge for a [YBCO$_{1.5}$/PBCO$_{5}$]
superlattice in cross section geometry. White contrast corresponds to PBCO
layers, while YBCO layers appear dark (b) Zoom view of figure 6(a).
Interfacial steps of one unit cell are clearly observed.

\bigskip

Figure 7: Low angle XRD spectra for a set of [YBCO$_{n}$/PBCO$_{5}$]
superlattices with noninteger YBCO layers. From bottom to top n=1.2, 1.5,
and 2 YBCO unit cells. Note the splitting (marked with dotted lines) between
the low angle satellite peaks and the satellites of the first Bragg peak
resulting of the incommensurate modulation.

\bigskip

Figure 8: Atomic force microscopy (AFM) image of a thick (300 \AA ) YBCO
film on a 5 u.c. PBCO buffer, showing a distribution of pyramid structures.
Terraces are 250 nm wide. Steps between terraces of the height of one unit
cell (11.7 \AA ) are observed.

\bigskip

Figure 9: c-lattice parameter (a) and critical temperature (b) as a function
of YBCO thickness for films grown on STO (100). Lines are guides for the eye.

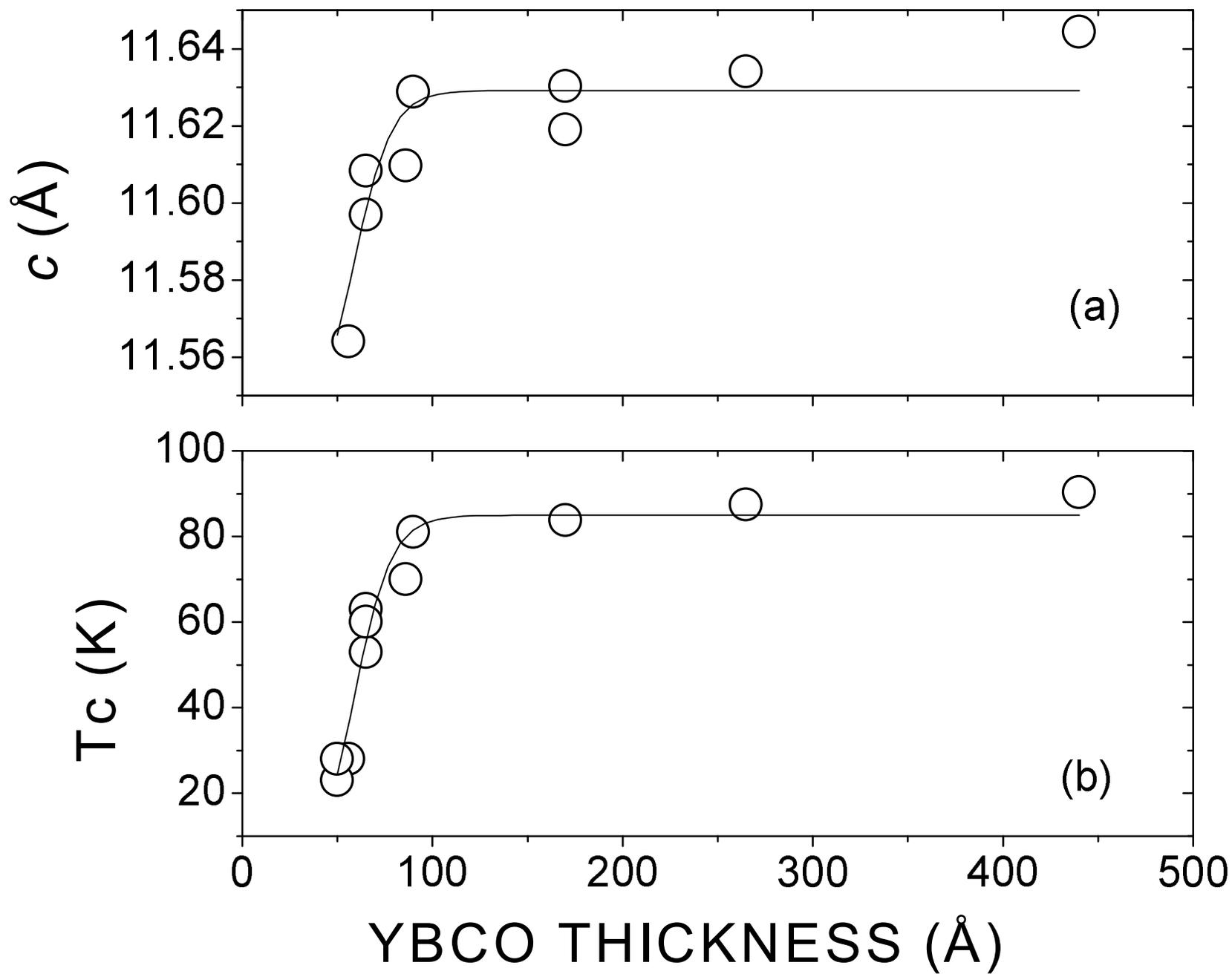
\end{document}